\documentclass[a4paper,10pt]{article}
\usepackage{latexsym}
\usepackage{amsmath}
\usepackage{amssymb}
\usepackage{graphicx}
\usepackage{wrapfig}
\usepackage{fancybox}
\usepackage{bm}
\title{Gauge Parameter Dependence of the 1-loop Fermion Self Energy at Finite Temperature}
\author{S.Sasagawa and H.Tanaka \\\\Department of Physics, Rikkyo University, Tokyo 171-8501, Japan}
\date{}
\begin{document}
\maketitle

\vspace{1em}
\vspace{1em}
\vspace{1em}
\begin{abstract}

We show a gauge parameter dependence of the 1-loop fermion self energy at finite temperature before the analytic continuation. We also show a gauge parameter dependence of only the temperature dependence term. The result is the same as the ladder approximation Schwinger-Dyson equation. The wave function renormalization constant approaches 1 by taking a negative gauge parameter. The analogy of the ladder approximation Schwinger-Dyson equation and the 1-loop calculation will help the analysis of the phase transition.

\end{abstract}

\vspace{1em}
\vspace{1em}
The fermion self energy after the analytic continuation is known as gauge invariance in the order $T^{2}$ by high temperature limit\cite{rf:1} or the hard thermal loop (HTL) approximation\cite{rf:11}. In addition, there are some topics beyond HTL approximation. For example, The numerical calculation without the HTL approximation\cite{rf:2}, the gauge dependence\cite{rf:3} and the 2-loop calculation\cite{rf:4} beyond the leading order HTL approximation.

On the other hand, the calculation of a critical point for a phase transition uses the self energy obtained by the Schwinger-Dyson equation (SDE) in the imaginary time formalism. (The critical point is obtained by inserting the self energy in the effective potential\cite{rf:5}.) However, the critical point obtained by using the ladder approximation has a gauge parameter dependence. Hence, in \cite{rf:6}, we showed the method with a gauge parameter that depends on an external momentum. Using this method, we performed the numerical calculation satisfying the Ward-Takahashi identity (WTI). Then, we obtained the critical point satisfying the WTI. However, we calculated without splitting the SDE into the divergent term and the term that should converge. Due to the structure of the SDE, it is difficult to calculate those separately. (We may expect that the divergent term and convergent term are split in the SDE, because a loop calculation at finite temperature is that way\cite{rf:11}. However, this expectation is uncertain.) Thus, we do not know a gauge parameter dependence of the convergent term. (Only the divergent term might be the same as the case of the zero temperature SDE.) However, the convergent term should take a major role in the phase transition. Hence, it is important to understand a gauge parameter dependence of the convergent term.

In this short article, to better understand a gauge parameter dependence of the SDE in the imaginary time formalism, we show a gauge parameter dependence of the 1-loop self energy before the analytic continuation.

The 1-loop massless fermion self energy in the finite temperature QED is given by

\begin{equation}
\displaystyle \begin{split}\Sigma(p)=&\displaystyle \frac{\alpha}{2\pi^{2}}T\sum_{l}\int d^{3}k\gamma^{\mu}\frac{g_{\mu\nu}}{(p-k)_{\alpha}\gamma^{\alpha}}\frac{1}{k^{2}}\gamma^{\nu}\\&+(\displaystyle \xi-1)\frac{\alpha}{2\pi^{2}}T\sum_{l}\int d^{3}k\gamma^{\mu}\frac{1}{(p-k)_{\alpha}\gamma^{\alpha}}\frac{1}{k^{2}}\frac{k_{\mu}k_{\nu}}{k^{2}}\gamma^{\nu},\end{split}\label{eq:1}
\end{equation}

\noindent where $p=(p_{0}=i\omega_{n},\mbox{\boldmath $p$}),\ k=(k_{0}=i\omega_{l},\mbox{\boldmath $k$}),\ \alpha=e^{2}/4\pi,\ \xi$ is a gauge parameter, $\omega_{n}$ and $\omega_{l}$ are the Matsubara frequency for fermions and bosons, respectively. Although we write as $\Sigma(p)$ for simplicity, a self energy or a propagator (Matsubara green function) depend $p_{0}$ and $|\mbox{\boldmath $p$}| $independently. The exact fermion propagator is defined as

\vspace{1em}
\begin{equation}
G(p)=\displaystyle \frac{-1}{p\hspace{-.50em}/-\Sigma(p)}.\\[0.28cm]
\end{equation}

\vspace{1em}
\noindent The general form of the fermion self energy at finite temperature is written by\cite{rf:1,rf:11}

\vspace{1em}
\begin{equation}
\Sigma(p)=-a(p)p_{0}\gamma_{0}+b(p)(\mbox{\boldmath $p$}\cdot\mbox{\boldmath $\gamma$}),\\[0.28cm]
\end{equation}

\noindent where $a(p)$ and $b(p)$ are given by

\vspace{1em}
\begin{equation}
a(p)=-\displaystyle \frac{1}{4p_{0}}\mathrm{tr}[\gamma_{0}\Sigma(p)]\ ,\ b(p)=-\frac{1}{4|\mbox{\boldmath $p$}|^{2}}\mathrm{t}\mathrm{r}[(\mbox{\boldmath $p$}\cdot\mbox{\boldmath $\gamma$})\Sigma(p)].\\[0.28cm]
\end{equation}

\noindent$a(p)$ and $b(p)$ are shown Appendix\ A. We show a gauge parameter dependence of $a(p)$ and $b(p)$. Since $a(p)$ and $b(p)$ are not physical quantities, there is no problem with $a(p)$ and $b(p)$ having a gauge dependence essentially.

$a(p)$ and $b(p)$ have two terms, that is, the term corresponding to zero temperature and the term including distribution functions\cite{rf:11}. We call the former the divergent term $a_{0}(p),b_{0}(p)$, the latter the temperature dependent term $a_{T}(p),b_{T}(p)$. The divergent term also have a temperature dependence before the analytic continuation $ i\omega_{n}\rightarrow p_{0}^{\prime}+i\eta$, and the temperature dependent term indicates the convergent term.

The divergent term needs a regularization. On numerical calculation, it is a simple method to adopt the ultraviolet cutoff $\Lambda$. However, the dependence of $\Lambda$ appears for $T/\Lambda>0.3$. (This dependence also appears for the SDE.\cite{rf:7}) The cutoff dependence arises from the convergence of the temperature dependent term.

\begin{figure}[t]

\begin{tabular}{cc}

\begin{minipage}{0.47\hsize}

\begin{center}

\includegraphics[width=67mm]{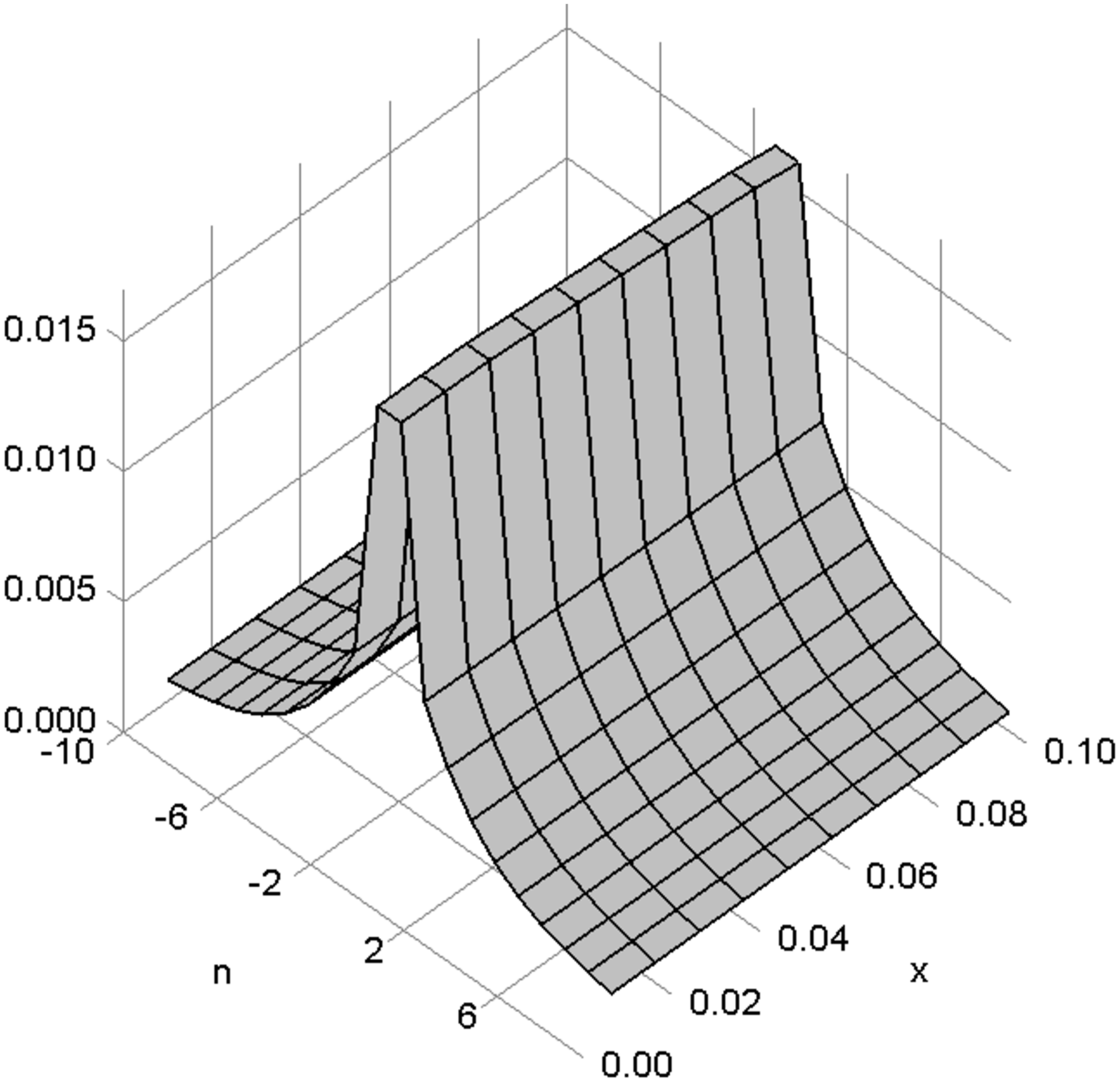}

\caption{The typical behavior of $a(p)$ with the Landau gauge at $T/\Lambda=0.1$ ($ x=|\mbox{\boldmath $p$}|/\Lambda$)}

\label{fig:3Da}

\end{center}

\end{minipage}

\hspace{0.1cm}

\begin{minipage}{0.47\hsize}

\begin{center}

\includegraphics[width=65mm]{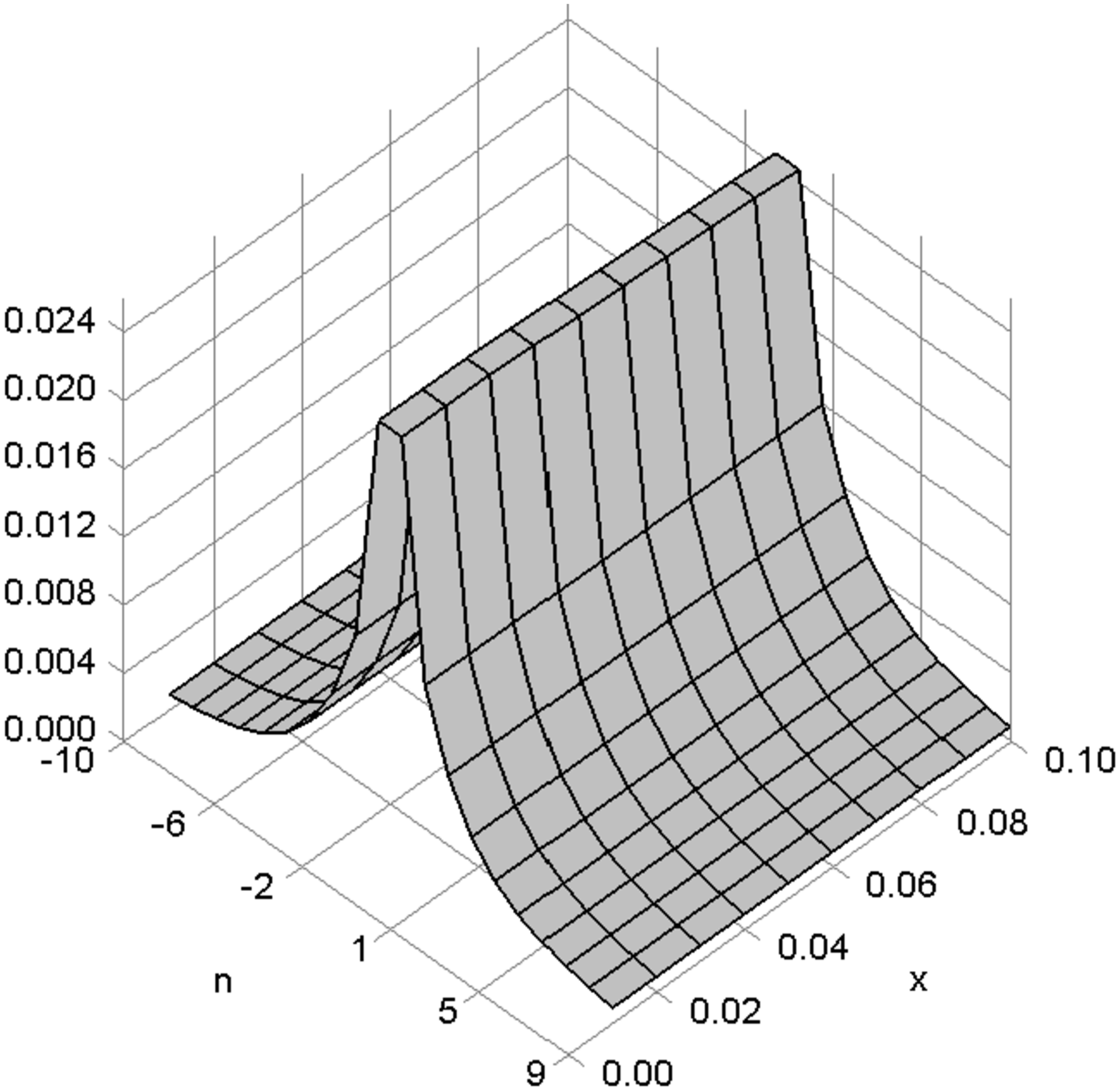}

\caption{The typical behavior of $b(p)$ with the Landau gauge at $T/\Lambda=0.1$ ($ x=|\mbox{\boldmath $p$}|/\Lambda$)}

\label{fig:3Db}

\end{center}

\end{minipage}

\end{tabular}

\end{figure}

\begin{figure}[t]

\begin{tabular}{cc}

\begin{minipage}{0.5\hsize}

\begin{center}

\includegraphics[width=65mm]{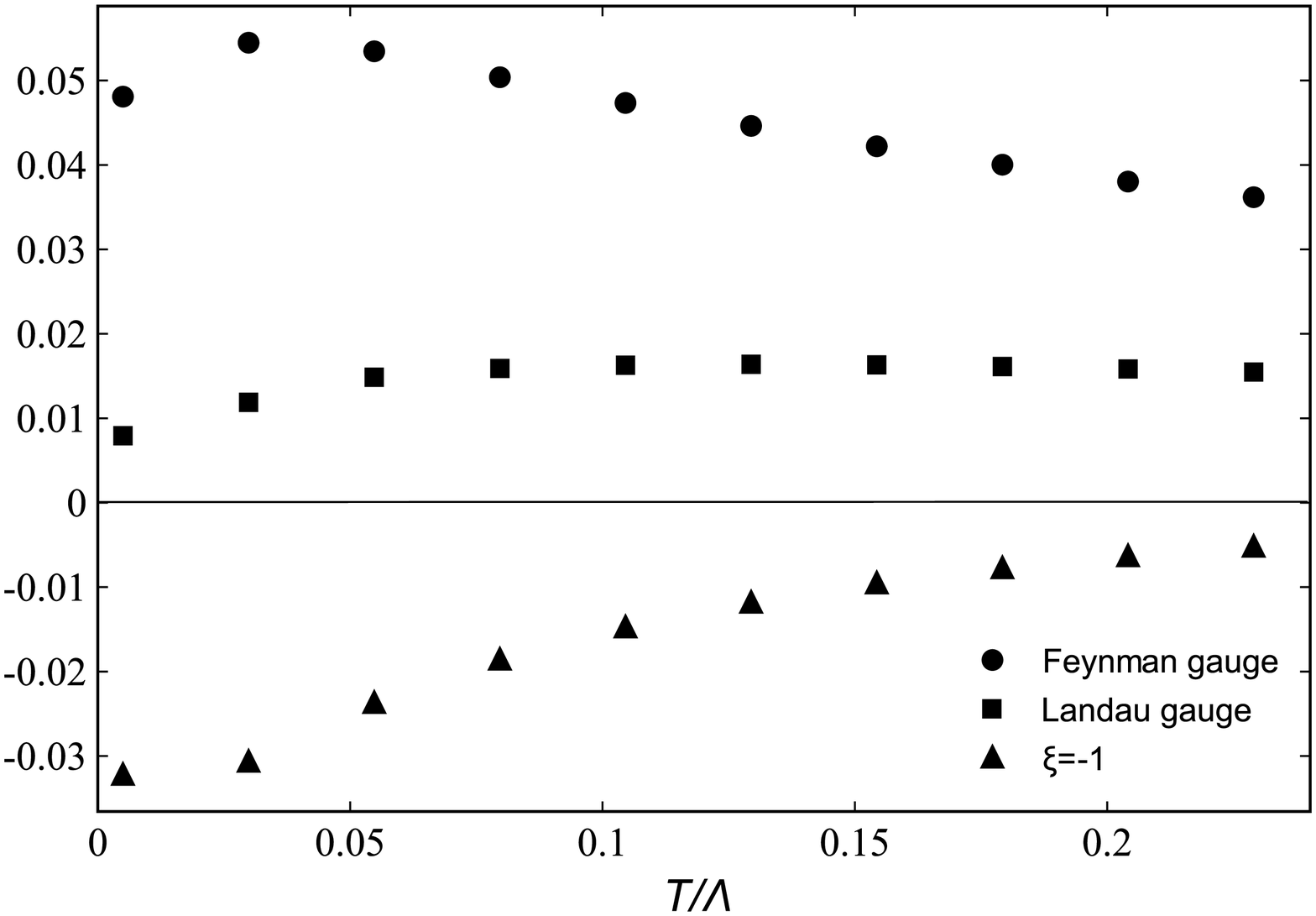}

\caption{$T$ dependence for $a(n=0,|\mbox{\boldmath $p$}|/\Lambda=0.1)$}

\label{fig:at0}

\end{center}

\end{minipage}

\begin{minipage}{0.5\hsize}

\begin{center}

\includegraphics[width=65mm]{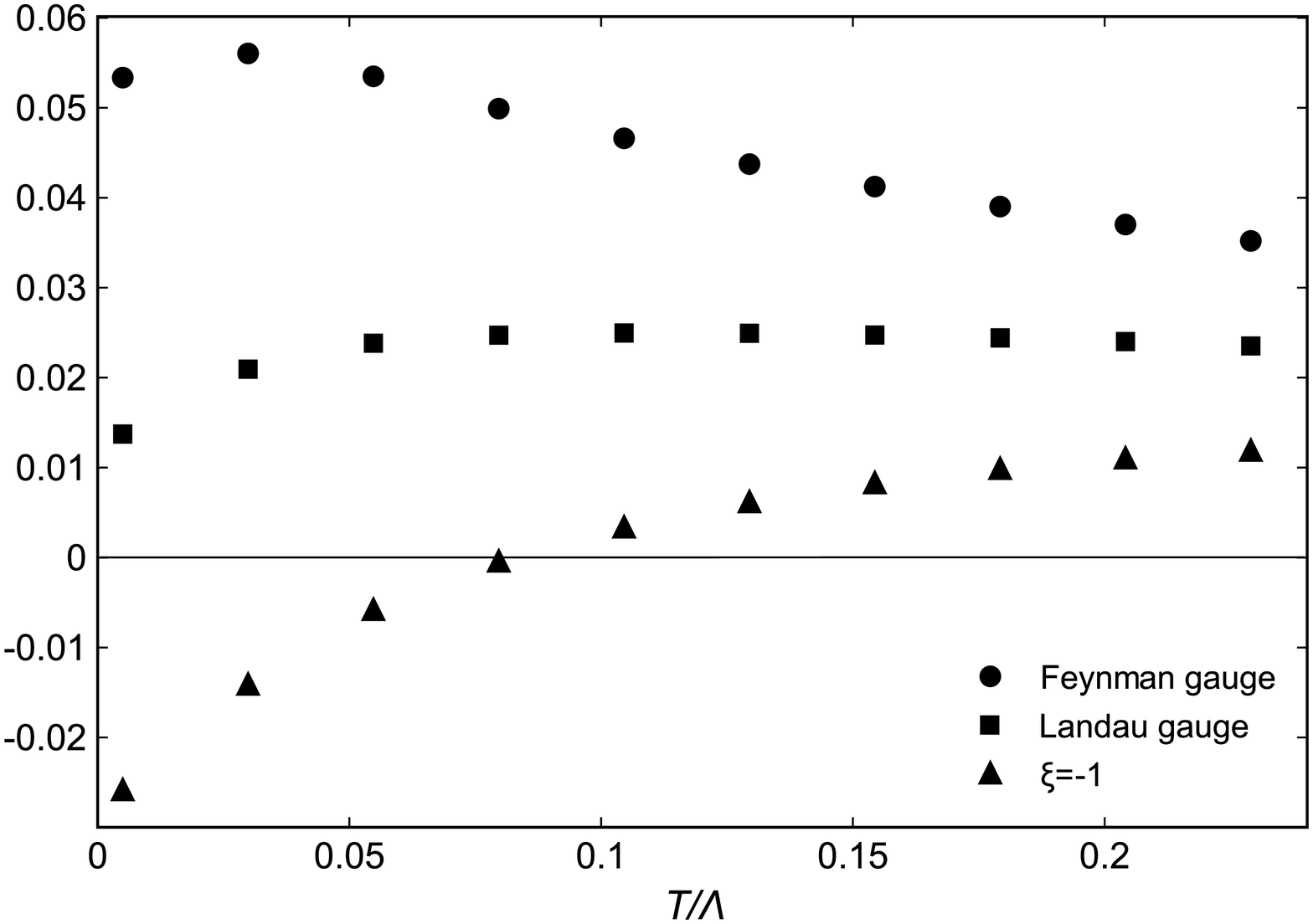}

\caption{$T$ dependence for $b(n=0,|\mbox{\boldmath $p$}|/\Lambda=0.1)$}

\label{fig:bt0}

\end{center}

\end{minipage}

\end{tabular}

\end{figure}

Results of the numerical calculation $a(p)$ and $b(p)$ are shown in Figs.\ \ref{fig:3Da}--\ref{fig:bt0}. We used $\alpha=0.1$ for all. Figs.\ \ref{fig:3Da} and \ref{fig:3Db} are a typical behavior for $n$ and $|\mbox{\boldmath $p$}|.\ a(p)$ and $b(p)$ for $n=-2\sim 1$ have large values, other values are small enough to ignore. Figs.\ \ref{fig:at0} and \ref{fig:bt0} show the temperature dependence of $a(p)$ and $b(p)$ for $n=0$ at various gauge parameters. Figs.\ \ref{fig:at0} and \ref{fig:bt0} correspond to \cite{rf:6}. In \cite{rf:6}, we showed $C_{n}(\mbox{\boldmath $p$})$ and $A_{n}(\mbox{\boldmath $p$})$ ($C_{n}(\mbox{\boldmath $p$})=1+a(p),A_{n}(\mbox{\boldmath $p$})=1+b(p)$) approach $1$ by moving a gauge parameter to a negative value. The same thing happened in the 1-loop calculation.

Taking the Landau gauge at zero temperature, the divergent term becomes zero in the 1-loop calculation or the ladder approximation SDE. On the other hand, $a(p)$ and $b(p)$ are not zero at finite temperature. One might say this arises from by the difference between the integral and the summation. (This is understood in clear term from the standard method for the summation\cite{rf:8}.) Thus, the difference is absorbed by moving a gauge parameter to a negative value.

\begin{figure}[t]

\begin{tabular}{cc}

\begin{minipage}{0.47\hsize}

\begin{center}

\includegraphics[width=62.5mm]{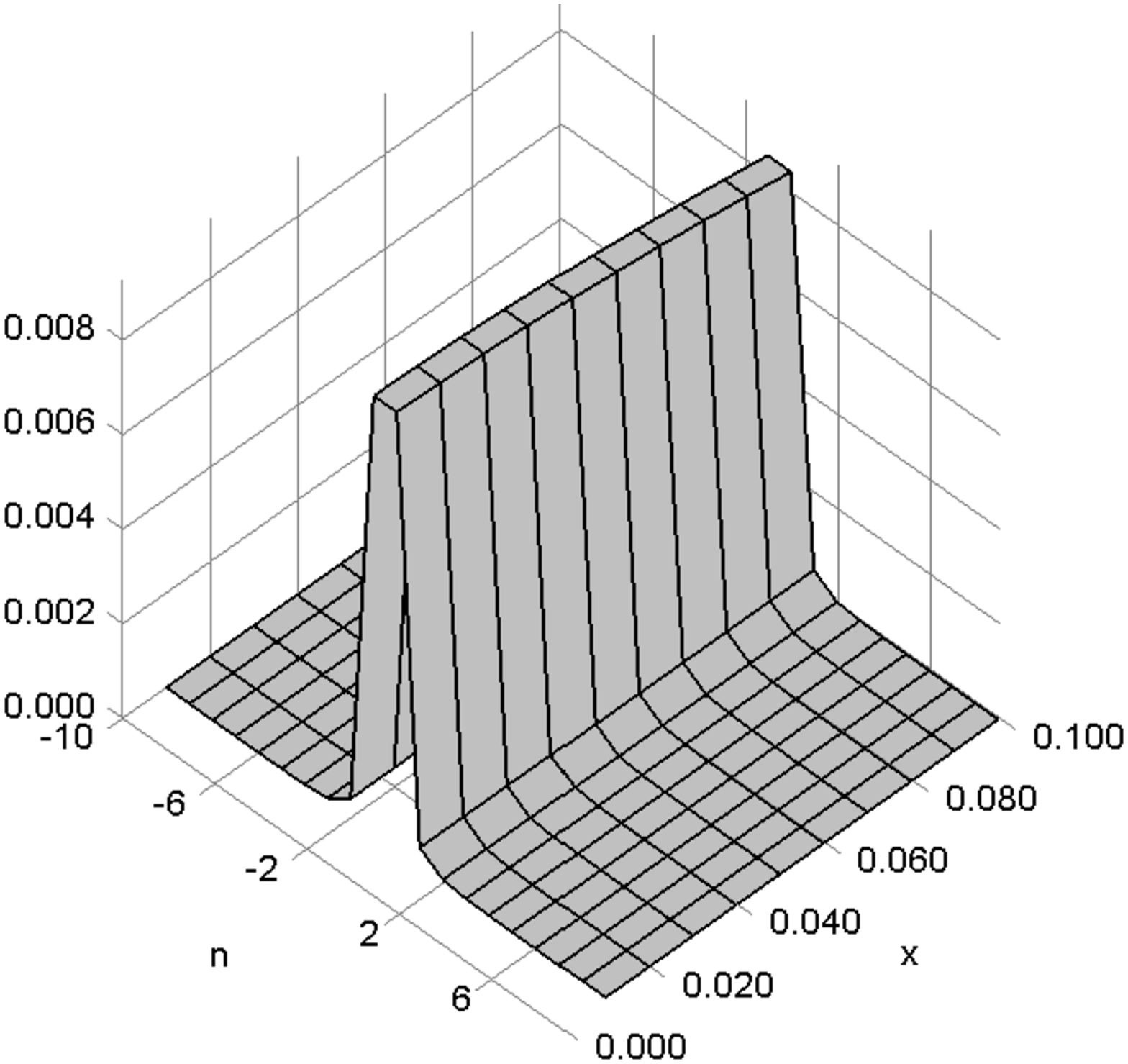}

\caption{The typical behavior of $a_{T}(p)$ with the Landau gauge at $T/\Lambda=0.1$ ($ x=|\mbox{\boldmath $p$}|/\Lambda$)}

\label{fig:3Dat}

\end{center}

\end{minipage}

\hspace{0.1cm}

\begin{minipage}{0.47\hsize}

\begin{center}

\includegraphics[width=62mm]{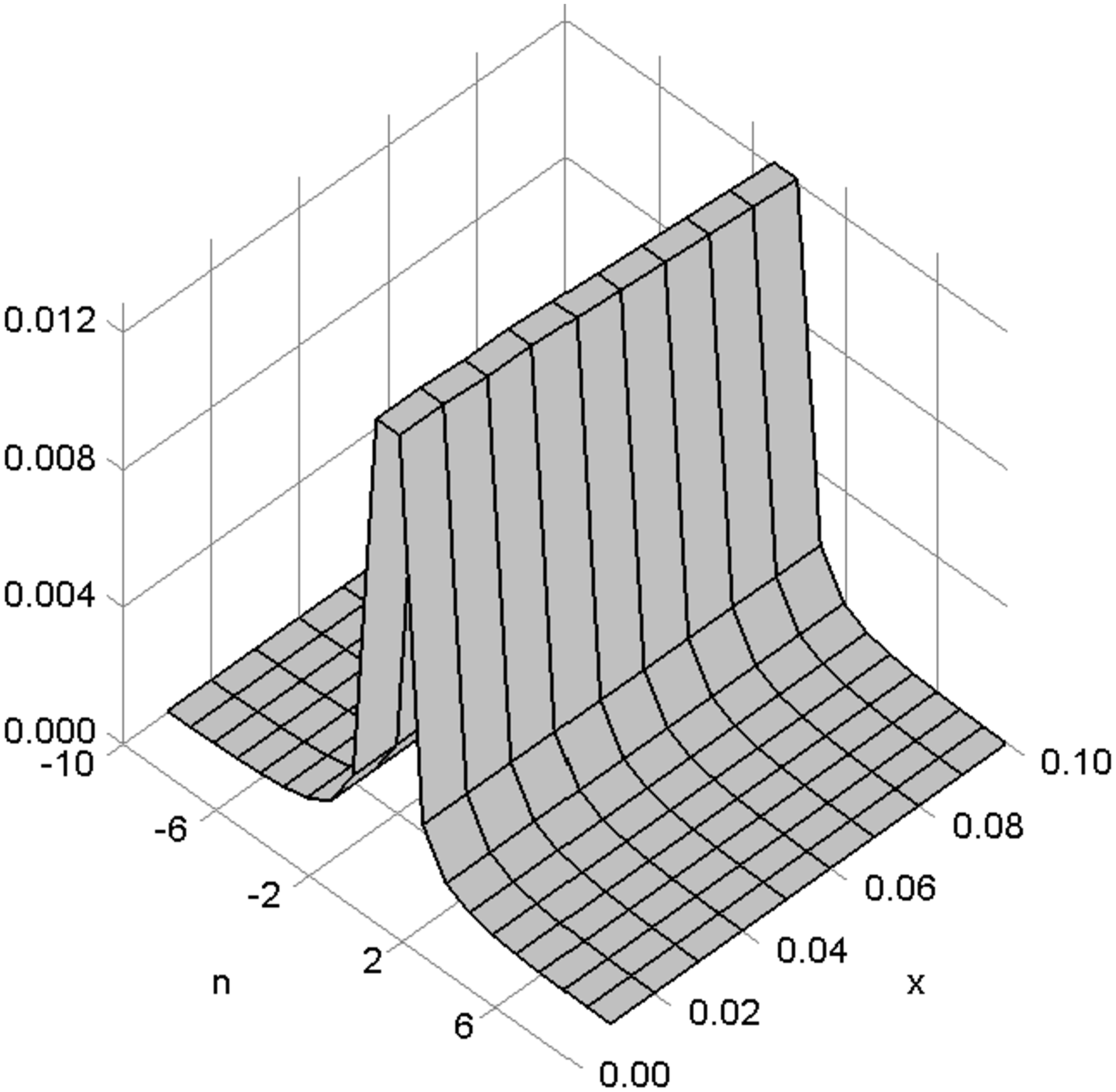}

\caption{The typical behavior of $b_{T}(p)$ with the Landau gauge at $T/\Lambda=0.1$ ($ x=|\mbox{\boldmath $p$}|/\Lambda$)}

\label{fig:3Dbt}

\end{center}

\end{minipage}

\end{tabular}

\end{figure}

\begin{figure}[t]

\begin{tabular}{cc}

\begin{minipage}{0.48\hsize}

\begin{center}

\includegraphics[width=65mm]{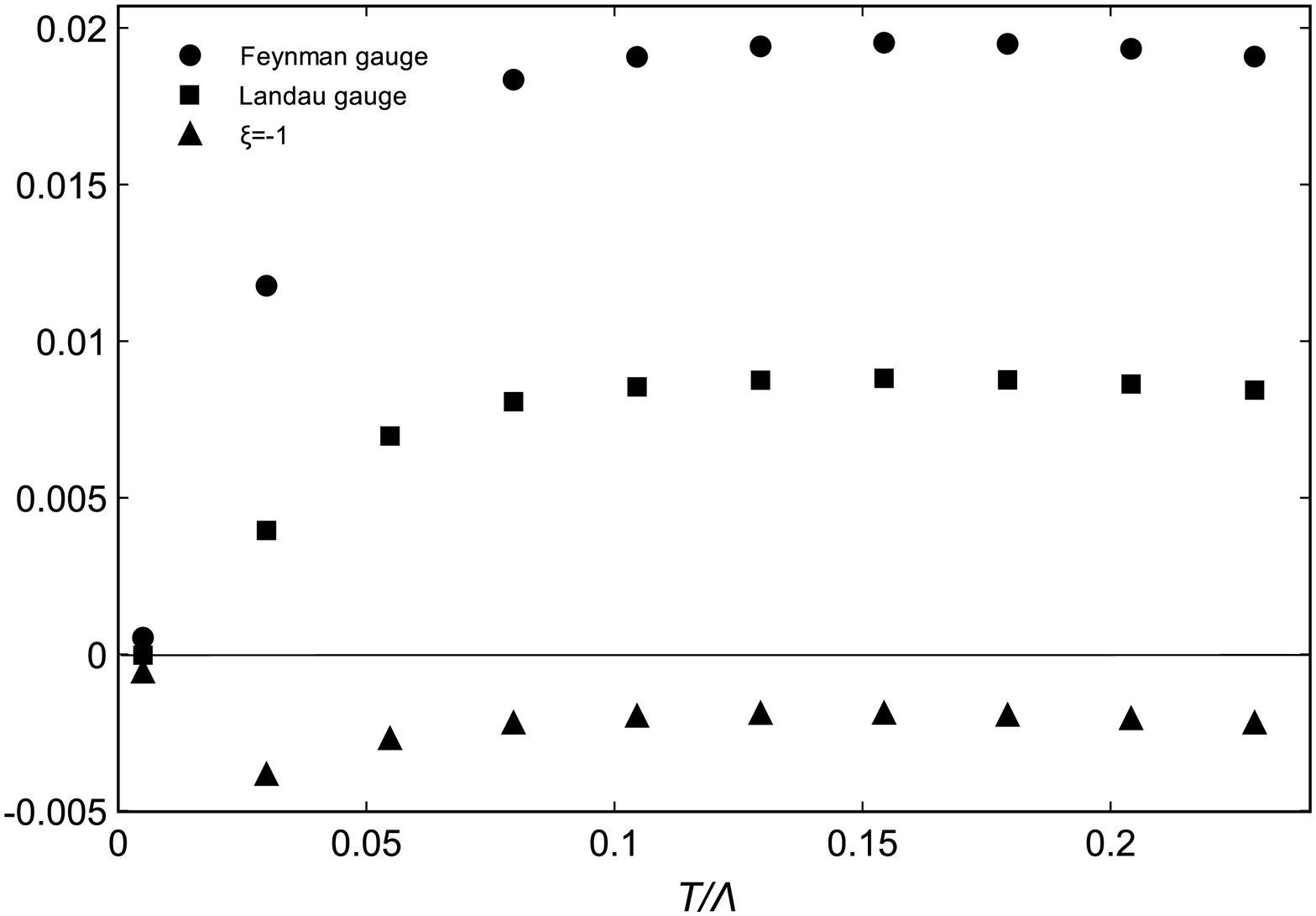}

\caption{$T$ dependence for $a_{T}(n=0,|\mbox{\boldmath $p$}|/\Lambda=0.1)$}

\label{fig:att}

\end{center}

\end{minipage}

\hspace{0.1cm}

\begin{minipage}{0.48\hsize}

\begin{center}

\includegraphics[width=65mm]{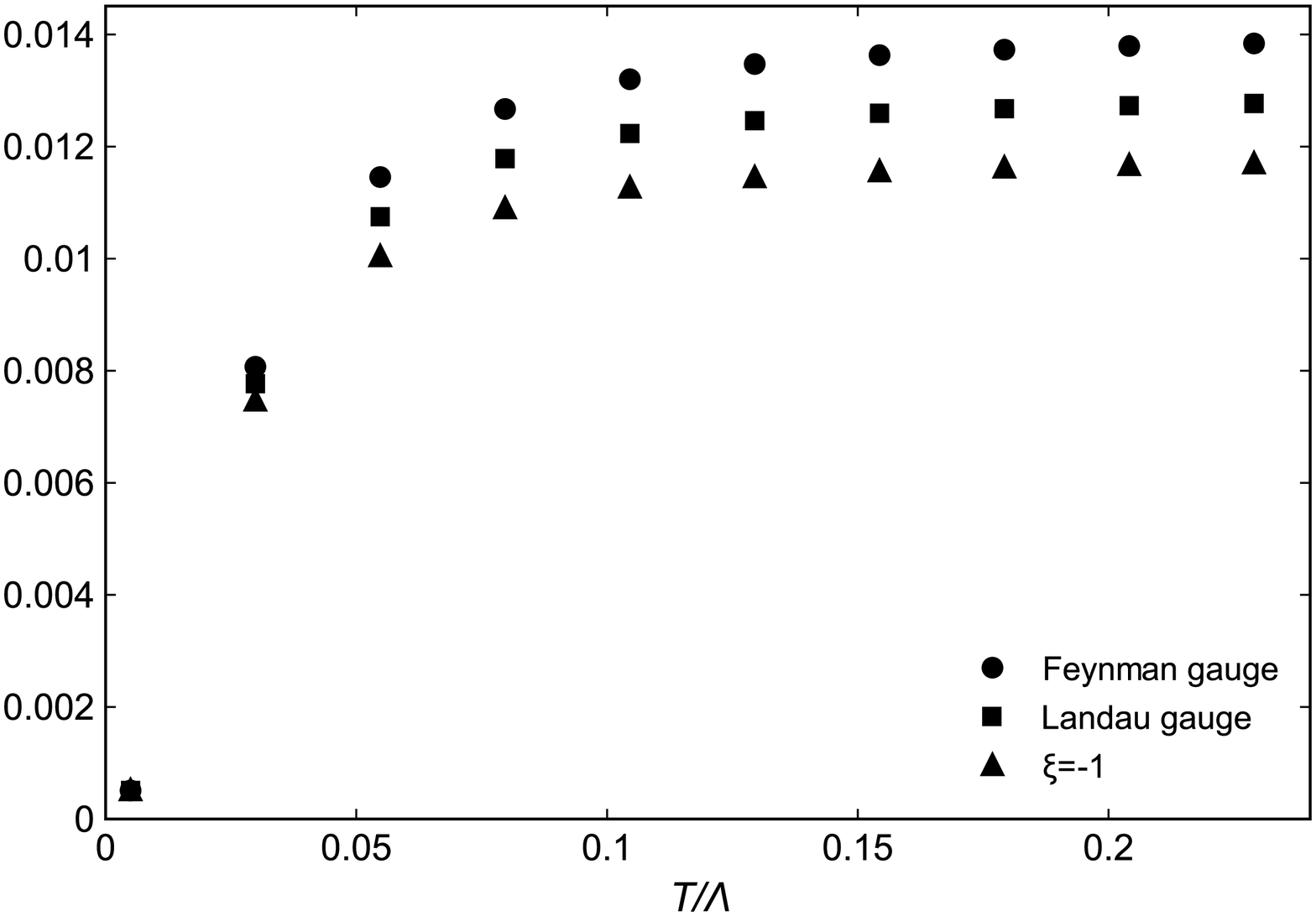}

\caption{$T$ dependence for $b_{T}(n=0,|\mbox{\boldmath $p$}|/\Lambda=0.1)$}

\label{fig:btt}

\end{center}

\end{minipage}

\end{tabular}

\end{figure}

Next, we show results of only the temperature dependent term. The temperature dependent term has the same gauge parameter dependence. The numerical result of the temperature dependent term $a_{T}(p)$ and $b_{T}(p)$ is shown Figs.\ \ref{fig:3Dat}--\ref{fig:btt}. The typical behavior of $a_{T}(p)$ and $b_{T}(p)$ is the same as $a(p)$ and $b(p)$. Although the cutoff $\Lambda$ is unnecessary for the temperature dependent term (because these converge), we normalize parameters by the same cutoff $\Lambda$ used in the divergent term.

Changing a gauge parameter, $a_{T}(p)$ or $b_{T}(p)$ becomes zero, because $a_{T}(p)$ or $b_{T}(p)$ approaches a constant by increasing temperature ($T>|\mbox{\boldmath $p$}|$). (Those for $n\neq-1,0$ are small enough to ignore.) For example, one takes approximately $a_{T}(p)\simeq 0$ to adopt the gauge parameter $\xi=-1$. (This choice might be also valid for $T<|\mbox{\boldmath $p$}|$.) On the other hand, ignoring the second term in (\ref{eq:1}) (the gauge dependent term) is better for $b_{T}(p)$ than the choice of a gauge parameter. Since this term in $b_{T}(p)$ is very small compared with the Feynman gauge term (see Fig.\ \ref{fig:btt}), one can ignore this term approximately. Thus, taking the gauge parameter $\xi=-1$ and ignoring the gauge dependent term of $b_{T}(p)$ might be a good approximation method for simplicity. The property, which one can ignore the temperature dependent term approximately by the choice of a gauge parameter, might be able to be used for a calculation including the self energy. If it is permitted to assume $a_{T}(p),$\ $b_{T}(p)=0$ approximately, one can use the result at zero temperature.

Furthermore, if one assumes that a gauge parameter depends to an external momentum and a temperature, one can make $a_{T}(p)=b_{T}(p)$. This is similarly possible after the analytic continuation. However, since plasmino modes appear as the gauge invariant form in the leading order HTL approximation\cite{rf:11}, this assumption is no meaningful at least after the analytic continuation.

We expect that propeties shown here are common with the ladder approximation SDE. In fact, a gauge parameter dependence of the case including the divergent term and the temperature dependent term are similar to the case of the SDE. Therefore, only the temperature dependent term might also have the common property in the 1-loop and the SDE. However, this expectation is uncertain (see appendix B). In addition, a temperature dependence of $a(p)$ and $b(p)$ in the strong QED is obviously different from the perturbative 1-loop calculation due to the chiral phase transition. To study this detail could help a understanding of the phase transition.

\vspace{1em}
\appendix

\section{$a(p)$ and $b(p)$}

When $\Sigma(p)\ $is defined as

\vspace{1em}
\begin{equation}
\Sigma(p)=\Sigma_{f}(p)+(\xi-1)\Sigma_{g}(p),
\end{equation}

\noindent($f$ expresses the term that adopting the Feynman gauge $\xi=1$ and $g$ expresses the gauge parameter term), corresponding terms $a_{f}(p), b_{f}(p), a_{g}(p),$ and $b_{g}(p)$ are written by

\vspace{1em}
\begin{align}
&a_{f}(p)=\displaystyle \frac{2e^{2}T}{p_{0}}\sum_{l}\int\frac{d^{3}k}{(2\pi)^{3}}\Big(\frac{p_{0}}{(p-k)^{2}k^{2}}+\frac{-k_{0}}{(p-k)^{2}k^{2}}\Big)\nonumber\\[0.2cm]
\vspace{1em}
&\hspace{2.2em}=2I_{1}-\displaystyle \frac{2}{p_{0}}I_{2},\\[0.25cm]
\vspace{1em}
&b_{f}(p)=2e^{2}T\displaystyle \sum_{l}\int\frac{d^{3}k}{(2\pi)^{3}}\Big(\frac{1}{(p-k)^{2}k^{2}}+\frac{1}{|\bm{p}|^{2}}\frac{-(\bm{p}\cdot \bm{k})}{(p-k)^{2}k^{2}}\Big)\nonumber\\[0.2cm]
\vspace{1em}
&\hspace{2.1em}=2I_{1}-\displaystyle \frac{2}{|\bm{p}|^{2}}I_{3},\\[0.25cm]
\vspace{1em}
&a_{g}(p)=\displaystyle \frac{e^{2}T}{p_{0}}\sum_{l}\int\frac{d^{3}k}{(2\pi)^{3}}\Big(\frac{-p_{0}}{(p-k)^{2}k^{2}}+\frac{k_{0}}{(p-k)^{2}k^{2}}-\frac{2p_{0}|\bm{k}|^{2}}{(p-k)^{2}k^{4}}+\frac{2k_{0}(\bm{p}\cdot \bm{k})}{(p-k)^{2}k^{4}}\Big)\nonumber\\[0.2cm]
\vspace{1em}
&\hspace{2.2em}=-I_{1}+\displaystyle \frac{1}{p_{0}}I_{2}-2K_{1}+\frac{2}{p_{0}}K_{2},\\[0.25cm]
\vspace{1em}
&b_{g}(p)=e^{2}T\displaystyle \sum_{l}\int\frac{d^{3}k}{(2\pi)^{3}}\Big(\frac{1}{(p-k)^{2}k^{2}}+\frac{1}{|\bm{p}|^{2}}\frac{\bm{p}\cdot \bm{k}}{(p-k)^{2}k^{2}}\nonumber\\[0.2cm]
\vspace{1em}
&\hspace{15.2em}+\displaystyle \frac{1}{|\bm{p}|^{2}}\frac{-2p_{0}k_{0}(\bm{p}\cdot \bm{k})}{(p-k)^{2}k^{4}}+\frac{1}{|\bm{p}|^{2}}\frac{2(\bm{p}\cdot \bm{k})^{2}}{(p-k)^{2}k^{4}}\Big)\nonumber\\[0.2cm]
\vspace{1em}
&\hspace{2em}=I_{1}+\displaystyle \frac{1}{|\bm{p}|^{2}}I_{3}-\frac{2p_{0}}{|\bm{p}|^{2}}K_{2}+\frac{2}{|\bm{p}|^{2}}K_{3},
\end{align}

\noindent After performing the summations and the angular integrals,

\vspace{1em}
\begin{align}
&I_{1}=-e^{2}\displaystyle \int_{0}^{\infty}\frac{d|\bm{k}||\bm{k}|^{2}}{(2\pi)^{3}}\frac{1}{2|\bm{k}|}\Big[n_{B}(|\bm{k}|)-n_{F}(|\bm{k}|)\Big]A_{1}+I_{1}(0),\\[0.22cm]
\vspace{1em}
&I_{2}=-e^{2}\displaystyle \int_{0}^{\infty}\frac{d|\bm{k}||\bm{k}|^{2}}{(2\pi)^{3}}\Big[\frac{1}{2}n_{B}(|\bm{k}|)A_{2}-\frac{1}{2|\bm{k}|}n_{F}(|\bm{k}|)A_{3}\Big]+I_{2}(0),\\[0.22cm]
\vspace{1em}
&I_{3}=-e^{2}\displaystyle \int_{0}^{\infty}\frac{d|\bm{k}||\bm{k}|^{2}}{(2\pi)^{3}}\Big[\frac{1}{2|\bm{k}|}(n_{B}(|\bm{k}|)+n_{F}(|\bm{k}|))A_{1}^{\prime}-\frac{|\bm{p}|^{2}}{2|\bm{k}|}n_{F}(|\bm{k}|)A_{1}\Big]+I_{3}(0),\\[0.22cm]
\vspace{1em}
&K_{1}=e^{2}\displaystyle \int_{0}^{\infty}\frac{d|\bm{k}||\bm{k}|^{2}}{(2\pi)^{3}}\Big[\frac{1}{4}\Big(\frac{n_{B}(|\bm{k}|)}{|\bm{k}|}-\frac{dn_{B}(|\bm{k}|)}{d|\bm{k}|}\Big)A_{1}-\frac{1}{2}n_{B}(|\bm{k}|)B_{4}+\frac{1}{2|\bm{k}|}n_{F}(|\bm{k}|)D\Big]+K_{1}(0),\\[0.22cm]
\vspace{1em}
&K_{2}=e^{2}\displaystyle \int_{0}^{\infty}\frac{d|\bm{k}||\bm{k}|^{2}}{(2\pi)^{3}}\Big[-\frac{1}{4|\bm{k}|}\frac{dn_{B}(|\bm{k}|)}{d|\bm{k}|}A_{2}^{\prime}-\frac{1}{2|\bm{k}|}\big(n_{B}(|\bm{k}|)+n_{F}(|\bm{k}|)\big)B_{3}^{\prime}+\frac{|\bm{p}|^{2}}{2|\bm{k}|}n_{F}(|\bm{k}|)B_{3}\Big]+K_{2}(0),\\[0.22cm]
\vspace{1em}
&K_{3}=e^{2}\displaystyle \int_{0}^{\infty}\frac{d|\bm{k}||\bm{k}|^{2}}{(2\pi)^{3}}\Big[\frac{1}{4|\bm{k}|^{2}}\Big(\frac{n_{B}(|\bm{k}|)}{|\bm{k}|}-\frac{dn_{B}(|\bm{k}|)}{d|\bm{k}|}\Big)A_{1}^{\prime\prime}-\frac{1}{2|\bm{k}|^{2}}n_{B}(|\bm{k}|)C_{4}\nonumber\\[0.22cm]
\vspace{1em}
&\hspace{10.2em}+\displaystyle \frac{1}{2|\bm{k}|}n_{F}(|\bm{k}|)\big(|\bm{p}|^{4}B_{1}+C_{1}-2|\bm{p}|^{2}B_{1}^{\prime}\big)\Big]+K_{3}(0),
\end{align}

\noindent where $I_{1}(0)\sim I_{3}(0),\ K_{1}(0)\sim K_{3}(0)$ are terms that does not include distribution functions, and distribution functions $n_{B}(|\bm{k}|) $and $n_{F}(|\bm{k}|)$ are

\vspace{1em}
\[
\frac{1}{e^{|\bm{k}|/T}-1},\ \frac{1}{e^{|\bm{k}|/T}+1},
\]

\noindent respectively. Other symbols are

\vspace{1em}
\begin{align}
&A_{1}=\displaystyle \frac{-\pi}{|\bm{p}||\bm{k}|}\log\Big|\frac{z_{+}^{2}+u^{2}}{z_{-}^{2}+u^{2}}\Big|,\\[0.22cm]
\vspace{1em}
&A_{2}=\displaystyle \frac{-2i\pi}{|\bm{p}||\bm{k}|}(\arctan\frac{z_{+}}{u}-\arctan\frac{z_{-}}{u}),\\[0.22cm]
\vspace{1em}
&A_{3}=\displaystyle \frac{-i\pi}{|\bm{p}||\bm{k}|}\Big[\omega_{n}\log\Big|\frac{z_{+}^{2}+u^{2}}{z_{-}+u^{2}}\Big|-2|\bm{k}|(\arctan\frac{z_{+}}{u}-\arctan\frac{z_{-}}{u})\Big],\\[0.22cm]
\vspace{1em}
&A_{1}^{\prime}=\displaystyle \frac{-\pi}{|\bm{p}||\bm{k}|}\Big[-4|\bm{p}||\bm{k}|-\frac{p^{2}}{2}\log\Big|\frac{z_{+}^{2}+u^{2}}{z_{-}^{2}+u^{2}}\Big|-u(\arctan\frac{z_{+}}{u}-\arctan\frac{z_{-}}{u})\Big],\\[0.22cm]
\vspace{1em}
&A_{2}^{\prime}=\displaystyle \frac{-i\pi}{|\bm{p}||\bm{k}|}\Big[\frac{u}{2}\log\Big|\frac{z_{+}^{2}+u^{2}}{z_{-}^{2}+u^{2}}\Big|-p^{2}(\arctan\frac{z_{+}}{u}-\arctan\frac{z_{-}}{u})\Big],\\[0.22cm]
\vspace{1em}
&A_{1}^{\prime\prime}=\displaystyle \frac{-\pi}{2|\bm{p}||\bm{k}|}\Big[4p^{2}|\bm{p}||\bm{k}|+\frac{1}{2}(p^{4}-u^{2})\log\Big|\frac{z_{+}^{2}+u^{2}}{z_{-}^{2}+u^{2}}\Big|+2p^{2}u(\arctan\frac{z_{+}}{u}-\arctan\frac{z_{-}}{u})\Big],\\[0.22cm]
\vspace{1em}
&B_{1}=\displaystyle \frac{2\pi}{|\bm{p}||\bm{k}|}\Big(\frac{z_{+}}{z_{+}^{2}+u^{2}}-\frac{z_{-}}{z_{-}^{2}+u^{2}}\Big),\\[0.22cm]
\vspace{1em}
&B_{3}=\displaystyle \frac{-2i\pi}{|\bm{p}||\bm{k}|}\Big[|\bm{k}|u\Big(\frac{1}{z_{+}^{2}+u^{2}}-\frac{1}{z_{-}^{2}+u^{2}}\Big)-\omega_{n}\Big(\frac{z_{+}}{z_{+}^{2}+u^{2}}-\frac{z_{-}}{z_{-}^{2}+u^{2}}\Big)\Big],\\[0.22cm]
\vspace{1em}
&B_{4}=\displaystyle \frac{-2\pi}{|\bm{p}||\bm{k}|}\Big[\omega_{n}u\Big(\frac{1}{z_{+}^{2}+u^{2}}-\frac{1}{z_{-}^{2}+u^{2}}\Big)+|\bm{k}|\Big(\frac{z_{+}}{z_{+}^{2}+u^{2}}-\frac{z_{-}}{z_{-}^{2}+u^{2}}\Big)\Big],\\[0.22cm]
\vspace{1em}
&B_{1}^{\prime}=\displaystyle \frac{-\pi}{|\bm{p}||\bm{k}|}\Big[u^{2}\Big(\frac{1}{z_{+}^{2}+u^{2}}-\frac{1}{z_{-}^{2}+u^{2}}\Big)+p^{2}\Big(\frac{z_{+}}{z_{+}^{2}+u^{2}}-\frac{z_{-}}{z_{-}^{2}+u^{2}}\Big)+\frac{1}{2}\log\Big|\frac{z_{+}^{2}+u^{2}}{z_{-}^{2}+u^{2}}\Big|\Big],\\[0.22cm]
\vspace{1em}
&B_{3}^{\prime}=\displaystyle \frac{-i\pi}{|\bm{p}||\bm{k}|}\Big[(\omega_{n}u^{2}-p^{2}|\bm{k}|u)\Big(\frac{1}{z_{+}^{2}+u^{2}}-\frac{1}{z_{-}^{2}+u^{2}}\Big)+(\omega_{n}p^{2}+|\bm{k}|u)\Big(\frac{z_{+}}{z_{+}^{2}+u^{2}}-\frac{z_{-}}{z_{-}^{2}+u^{2}}\Big)\nonumber\\[0.22cm]
\vspace{1em}
&\hspace{6.2em}+\displaystyle \frac{\omega_{n}}{2}\log\Big|\frac{z_{+}^{2}+u^{2}}{z_{-}^{2}+u^{2}}\Big|-i|\bm{k}|(\arctan\frac{z_{+}}{u}-\arctan\frac{z_{-}}{u})\Big],\\[0.22cm]
\vspace{1em}
&C_{1}=\displaystyle \frac{-\pi}{2|\bm{p}||\bm{k}|}\Big[-4|\bm{p}||\bm{k}|-2p^{2}u^{2}\Big(\frac{1}{z_{+}^{2}+u^{2}}-\frac{1}{z_{-}^{2}+u^{2}}\Big)+(-p^{4}+u^{2})\Big(\frac{z_{+}}{z_{+}^{2}+u^{2}}-\frac{z_{-}}{z_{-}^{2}+u^{2}}\Big)\nonumber\\[0.22cm]
\vspace{1em}
&\hspace{6.2em}-p^{2}\displaystyle \log\Big|\frac{z_{+}^{2}+u^{2}}{z_{-}^{2}+u^{2}}\Big|-2u(\arctan\frac{z_{+}}{u}-\arctan\frac{z_{-}}{u})\Big],\\[0.22cm]
\vspace{1em}
&C_{4}=\displaystyle \frac{-\pi}{2|\bm{p}||\bm{k}|}\Big[4|\bm{p}||\bm{k}|^{2}+(-\omega_{n}u^{3}+\omega_{n}p^{4}u+2p^{2}|\bm{k}|u^{2})\Big(\frac{1}{z_{+}^{2}+u^{2}}-\frac{1}{z_{-}^{2}+u^{2}}\Big)\nonumber\\[0.22cm]
\vspace{1em}
&\hspace{6.2em}+(-2p^{2}\displaystyle \omega_{n}u+p^{4}|\bm{k}|-|\bm{k}|u^{2})\Big(\frac{z_{+}}{z_{+}^{2}+u^{2}}-\frac{z_{-}}{z_{-}^{2}+u^{2}}\Big)+(-\omega_{n}u+p^{2}|\bm{k}|)\log\Big|\frac{z_{+}^{2}+u^{2}}{z_{-}^{2}+u^{2}}\Big|\nonumber\\[0.22cm]
\vspace{1em}
&\hspace{6.2em}+(2p^{2}\displaystyle \omega_{n}+2|\bm{k}|u)(\arctan\frac{z_{+}}{u}-\arctan\frac{z_{-}}{u})\Big],\\[0.22cm]
\vspace{1em}
&D=\displaystyle \frac{-2\pi}{|\bm{p}||\bm{k}|}\Big[-u^{2}\Big(\frac{1}{z_{+}^{2}+u^{2}}-\frac{1}{z_{-}^{2}+u^{2}}\Big)+(\omega_{n}^{2}-|\bm{k}|^{2})\Big(\frac{z_{+}}{z_{+}^{2}+u^{2}}-\frac{z_{-}}{z_{-}^{2}+u^{2}}\Big)-\frac{1}{2}\log\Big|\frac{z_{+}^{2}+u^{2}}{z_{-}^{2}+u^{2}}\Big|\Big],
\end{align}

\noindent where $z_{+}=p^{2}-2|\bm{p}||\bm{k}|,\ z_{-}=p^{2}+2|\bm{p}||\bm{k}|,$ and $u=2\omega_{n}|\bm{k}|$.

\vspace{1em}
\section{Separating of the SDE}

To give one example of the decomposition of the SDE, we use the ladder and instantaneous exchange (IE) approximation\cite{rf:13}. The IE approximation is given by

\vspace{1em}
\[
D_{\mu\nu}(k_{0},\bm{k})\ \Rightarrow\ D_{\mu\nu}(k_{0}=0,\bm{k}),
\]

\noindent where $D_{\mu\nu}(k_{0},\bm{k})$ is the gauge boson propagator. Since the fermion self energy becomes $p_{0}$ independent, we can perform the summation immediately. Thus, the summation in the SDE is

\vspace{1em}
\begin{align}
T\displaystyle \sum_{l}\frac{1}{q_{0}^{2}-M^{2}}=&-\displaystyle \frac{1}{2M}+\frac{1}{2M}\frac{1}{e^{M/T}+1}+\frac{1}{2M}\frac{1}{e^{M/T}+1}\nonumber\\[0.21cm]
\vspace{1em}
=&-\displaystyle \frac{1}{2M}\tanh(\frac{M}{2T}).
\end{align}

\noindent Using the exact fermion propagator,

\vspace{1em}
\[
G_{n}(\bm{p})=\frac{-1}{C_{n}(\bm{p})p_{0}\gamma_{0}+A_{n}(\bm{p})p^{i}\gamma_{i}-B_{n}(\bm{p})},
\]

\noindent the ladder and the IE approximation SDE at finite temperature is given by

\vspace{1em}
\begin{align}
&C(x)=1,\\[0.22cm]
\vspace{1em}
&A(x)=1+\displaystyle \frac{\alpha}{4\pi x^{3}}\int_{0}^{\infty}dy\ \frac{yA(y)}{M}\Big[-4xy+(x^{2}+y^{2})\log\frac{(x+y)^{2}}{(x-y)^{2}}\nonumber\\[0.22cm]
\vspace{1em}
&\hspace{13.9em}+\displaystyle \frac{\xi-1}{2}\Big(-4xy+(x^{2}+y^{2})\log\frac{(x+y)^{2}}{(x-y)^{2}}\Big)\Big]\tanh(\frac{M}{2T}),\label{eq:A1}\\[0.22cm]
\vspace{1em}
&B(x)=\displaystyle \frac{\alpha(3+\xi)}{4\pi x}\int_{0}^{\infty}dy\ \frac{yB(y)}{M}\log\frac{(x+y)^{2}}{(x-y)^{2}}\tanh(\frac{M}{2T}),\label{eq:B1}
\end{align}

\noindent where $M=\sqrt{y^{2}A^{2}(y)+B^{2}(y)},\ x=|\bm{p}|,$\ and $y=|\bm{q}|$. Due to $p_{0}$ independent, $C(x)$ must be $1$. Furthermore,\ adopting $\xi=-1,\ A(x)$\ is $1$. The gauge parameter dependence of $A(x)$ is distributed symmetrically.

Adopting $\xi=-1$, we consider only $B(x)$. We assume naively that the SDE is rewritten as

\vspace{1em}
\begin{equation}
B(x)=\displaystyle \frac{\alpha}{\pi x}\int_{0}^{\infty}dy\ \frac{yB_{0}(y)}{2M_{0}}\log\frac{(x+y)^{2}}{(x-y)^{2}}-\frac{\alpha}{\pi x}\int_{0}^{\infty}dy\ \frac{yB_{T}(y)}{M_{T}}\log\frac{(x+y)^{2}}{(x-y)^{2}}\frac{1}{e^{M_{T}/T}+1},\label{eq:B2}
\end{equation}

\noindent where $B(x)=B_{0}(x)-B_{T}(x)\ (B_{0}(x),B_{T}(x)\geq 0$, because $B(x)$ becomes zero$)$ and $M_{0,T} =\sqrt{y^{2}+B_{0,T}^{2}(y)}$. The first term is temperature independent, and the second term is temperature dependent. The temperature dependent term does not have a divergence.

\begin{figure}[t]

\begin{tabular}{cc}

\begin{minipage}{0.48\hsize}

\begin{center}

\includegraphics[width=60mm]{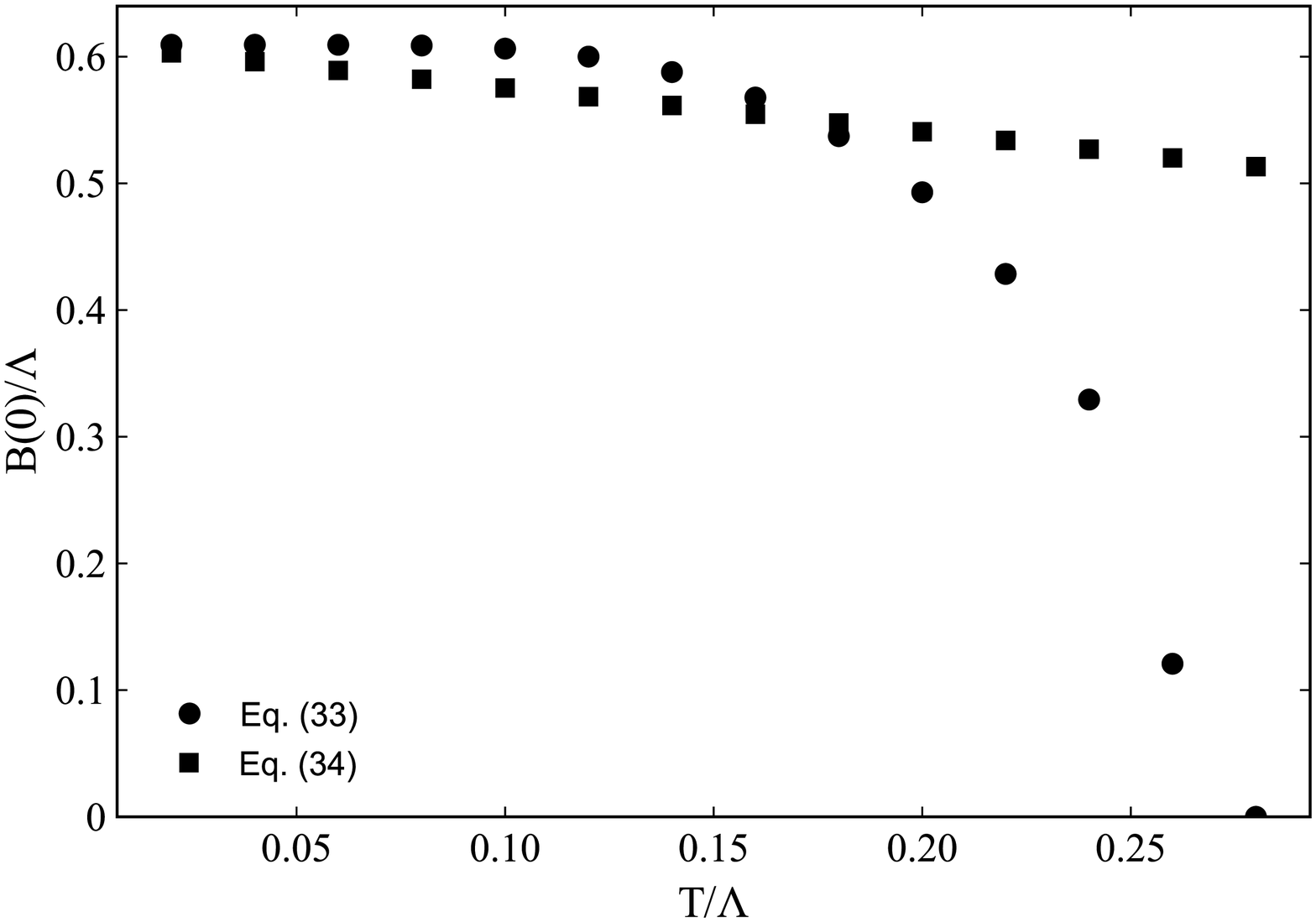}

\caption{The comparison of (\ref{eq:B1}) with (\ref{eq:B2}). $\alpha=1.5$.\newline}

\label{fig:B}

\end{center}

\end{minipage}

\hspace{0.1cm}

\begin{minipage}{0.48\hsize}

\begin{center}

\includegraphics[width=60mm]{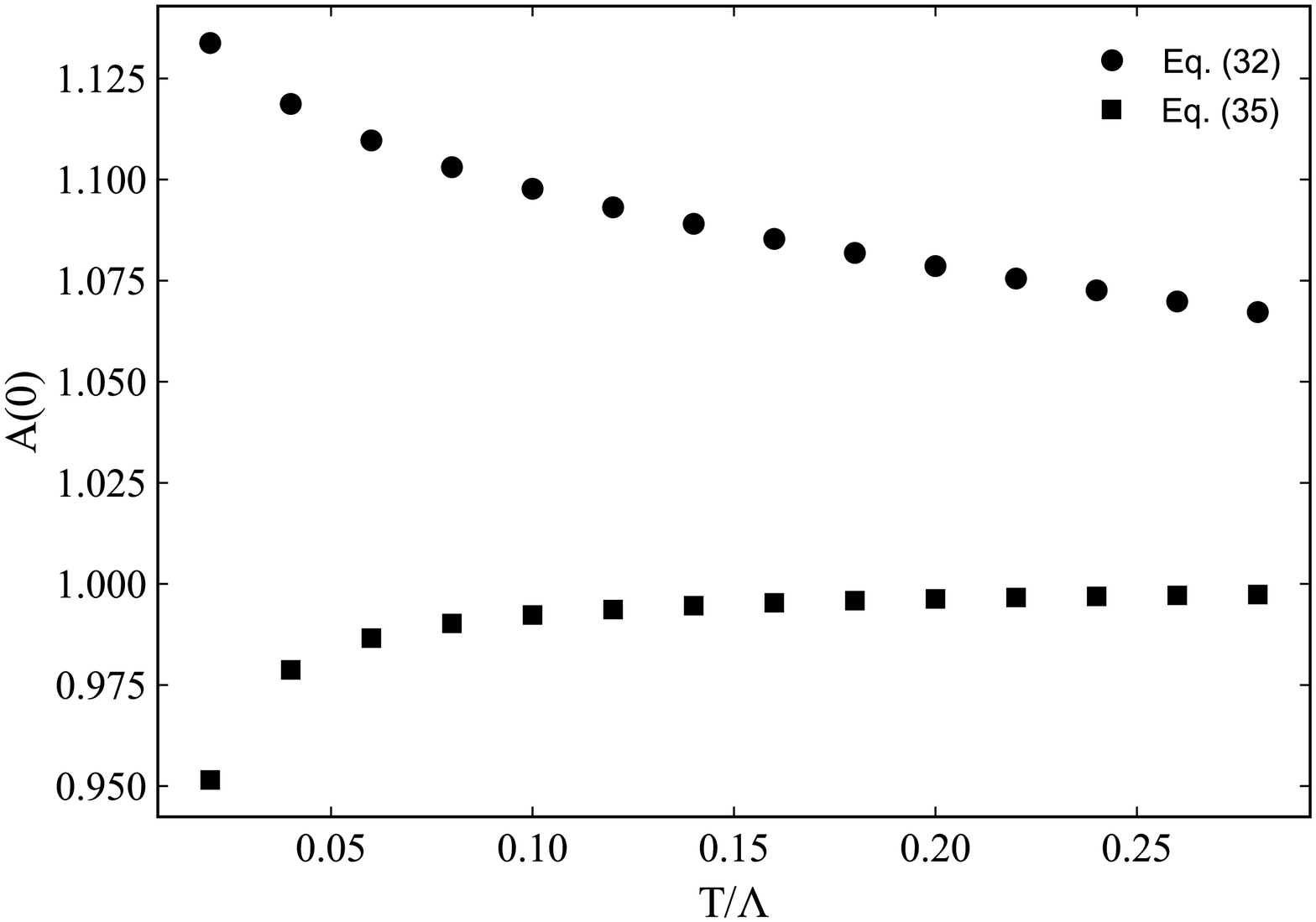}

\caption{The comparison of (\ref{eq:A1}) with (\ref{eq:A2}). $\alpha=0.1,\ \xi=0$.}

\label{fig:A}

\end{center}

\end{minipage}

\end{tabular}

\end{figure}

The numerical results for a temperature dependence are shown Fig.\ \ref{fig:B}. Since the temperature dependent term slowly decreases, the critical temperature is at a much higher temperature than the usual result. The value is totally impractical. Thus, this assumption is not available in the study of the chiral phase transition.

Using the same assumption, $A(x)$ in the chiral limit is rewritten as

\vspace{1em}
\begin{align}
A(x)=1+\displaystyle \frac{\alpha}{2\pi x^{3}}\int_{0}^{\infty}dy\Big[&-4xy+(x^{2}+y^{2})\displaystyle \log\frac{(x+y)^{2}}{(x-y)^{2}}\nonumber\\[0.22cm]
\vspace{1em}
&+\displaystyle \frac{\xi-1}{2}\Big(-4xy+(x^{2}+y^{2})\log\frac{(x+y)^{2}}{(x-y)^{2}}\Big)\Big]\Big(\frac{1}{2}-\frac{1}{e^{M_{A}/T}+1}\Big),\label{eq:A2}
\end{align}

\noindent where $A(x)=A_{0}(x)+A_{T}(x)$ and $M_{A}=yA_{T}(y)$. Fig.\ \ref{fig:A} shows that this assumption is not available, because $A_{T}(x)$ does not become zero in the zero temperature limit.

The SDE does not have a simple structure even allowing for the IE approximation. This difficulty should be produced by the divergent term (zero temperature like term) and the temperature dependent term.

\end{document}